\documentclass[11pt, preprint,superscriptaddress]{revtex4}
\usepackage{hyperref}
\usepackage{amsmath,amsfonts}
\usepackage[dvips]{graphics}
\usepackage[dvips]{graphicx}
\usepackage{subfigure}
\usepackage{epsfig}
\usepackage{amsmath}
\usepackage{amsfonts}
\RequirePackage{color}

\begin{document}
\title{\bf
Effect of electromagnetic permeability on transition temperature of superconductivity}

\author{M. M. Bagheri-Mohagheghi}
 \email{bmohagheghi@du.ac.ir}
  \author{B. Pourhassan}
  \email{b.pourhassan@du.ac.ir}
   \author{M. Adelifard}
    \affiliation{School of Physics, Damghan University, P. O. Box 3671641167, Damghan, Iran}
       \author{M. S.-Saremi}
     \affiliation{Department of Electrical Engineering, Ferdowsi University of Mashhad, Mashhad, Iran}
         \author{S. Upadhyay}
  \email{sudhakerupadhyay@gmail.com; sudhaker@associates.iucaa.in}
   \affiliation{Department of Physics, K. L. S. College, Nawada, Bihar 805110, India}
\affiliation{Department of Physics, Magadh University, Bodh Gaya,
 Bihar  824234, India}
\affiliation{Inter-University Centre for Astronomy and Astrophysics (IUCAA), Pune, \\Maharashtra 411007, India}

\begin{abstract}
The description of superconductivity at high temperature is a problem that has recently been addressed. Transition temperature of superconductivity, $T_c$, depends on the lattice structure type, size, and room pressure. In super-lattices and low dimensional layered nanostructures, $T_c$ is increased, by increasing the complexity of the structure and internal pressures in solid lattice. In this paper, we investigate the relation between physical parameters ($\epsilon, \mu$) of matter and superconductivity properties as well as transition temperature ($T_c$), and explain the superconductivity at high temperature. In this study, a semi-classical electromagnetic description along with vortex topologic theory and quantum dynamic models with experimental data considered to justify the relation between superconductivity phenomena and magnetic monopole properties. We find that the electromagnetic energy of magnetic monopole is in agreement with vortex energy in topological theory and it can get close to thermal energy at high temperature. These models suggest the superconductivity is related to the mobile monopole or vortices. we show that the electrical permittivity ($\epsilon$) and magnetic permeability ($\mu$) of matter have a key role in the superconductive properties.\\
{\bf Keywords:} Superconductivity; Magnetic properties; Electrical permittivity; Magnetic permeability.
\end{abstract}
\maketitle
\section{Introduction}\label{sec1}
 Superconductivity is one of the important subjects in theoretical and experimental physics with one century historical background. For the first time, the superconductor phase transition was discovered in 1911 by K. Onnes. Superconductivity is a reversible thermodynamic state of material in which resistivity is disappeared ($\rho = 0$). It occurs below a critical temperature ($T_c$), current level ($J_c$), and magnetic field ($H_c$). The superconductors are considered as perfect diamagnetic materials, so that when it is placed in a magnetic field $B$, the flux lines avoid the region of magnetic flow.  It is a fundamental property of superconductors and for the first time explained by Meissner in 1933 which is known as the Meissner effect \cite{1,2,3,4,5}.

Some typical critical temperatures for elemental superconductors are as follows: $Nb$ at $9.25 K$, $Pb$ at $7.2 K$, $Al$ at $1.1 K$, and silicon at $7.1 K$ (under high pressure). However, noble metals have not yet been found to have a superconductor phase transition even at very low temperatures of a few milli-Kelvins. The resistance of a conventional metal decreases gradually by decreasing temperature to very low temperatures, and may be proportional to fifth power of temperature ($\rho  \propto  T^5$) while a superconductor shows zero resistivity below $T_c$ \cite{6}. The first high-$T_c$ superconductor was discovered in 1986 by IBM researchers,  who were awarded the 1987 Nobel prize in physics for their attempts to the discovery of superconductivity in ceramic materials \cite{4}.

The usual way to study the low temperature superconductors is BCS theory. For high temperature superconductors, it is indicated that it is in strong coupled regime, so a departure from the quasi-particle projection of the Fermi liquid theory is needed \cite{7}. In that case, there is a few comprehensive method to study high temperature superconductor in condense matter physics. Recently, by B. C. Sinha and J. Apoorv, present a combined mechanism involving phononic and electronic processes is suggested for high Tc-superconductivity in doped graphene. It is found that on increasing the doping such as B and Al in graphene, the superconducting critical temperature can be raised to room temperature \cite{8}.  However, AdS/CFT correspondence  is a powerful tool to study superconductors in high temperature
 \cite{9,10}. Recently, a model of holographic superconductor has been presented and found that magnetic monopole enhances thermal and electrical properties of superconductivity \cite{11}.  On the other hand, a classical study of monopoles can be found in \cite{12} and also in \cite{13}, superconductivity in magnetic multipole states have been studied. Thereinafter, the model of magnetic monopoles has been widely used to describe many experimental observations in spin ice materials such as $Dy_2Ti_2O_7$ and $Dy_2Sn_{2}O_7$ \cite{14,15,16,17,18}.

Newly, a topological model for study of critical phenomena and phase transitions by "Thouless, Kosterlitz and Haldane (TKH)" (winners of physics Nobel prize 2016) are established.  In this theory, phase transitions of matter are related to the topological defects i.e. vortices. By this model, the phase change in critical phenomena such as superconductivity, magnetic ultra-thin films, and superfluid by topological defects are explainable \cite{19,20}.
In this paper, by using semi-classical electrodynamics calculations and quantum mechanic, we show that magnetic monopole can exist at low temperatures up to room temperature and superconductivity with origin of the magnetic monopole can exist at critical temperature range of  $T_c \leq 313K$, depended on dielectric constant and lattice structure. This model is in agreement with advanced topological phase-transition model of TKH in superconductors. Finally, we conclude that origin of superconductivity is magnetism.

The paper is organized as follows: In the section 2 we give brief review of relation between $T_c$ superconductivity temperature and energy magnetic monopoles. In section 3-1 a
 semi-classical model to justify the existence of magnetic monopole in superconductive media is presented. In section 3-2 a topological model for magnetic monopoles similar to vortex in superconductive matter is presented. In section 3-3 a simple quantum mechanics model for calculating the energy of magnetic monopole is introduced. In section 3-4, we explain the dynamic motion of magnetic monopoles or vortex and its relation with electrical conductivity. Then, in discussion section, we put forward explanations and arguments about the relation between magnetic monopoles and superconductivity phenomenon with structural, electro-magnetic and topological views. Especially, with presenting the examining parameters such as dielectric constant and lattice structure, possibility of monopole creation in high pressure and lattice stress conditions is investigated. Finally, we give conclusion and proposals for future researches.

\section{Scientific reports on Superconductor and magnetic monopole}\label{sec2}
Some typical critical temperatures of important superconductors are given in TABLE \ref{tb1}. The structure of high-$T_c$ copper oxide or cuprite superconductors are often closely related to perovskite structure, and the structure of these compounds has been described as a distorted, oxygen deficient multi-layered perovskite structure \cite{4}.
 \begin{table}[ht]
\caption{Transition temperature
(in Kelvin)} \vspace{.3cm} 
\centering 
\begin{tabular}{c c c } 
 \hline\hline 
$T_c$  \ \  &   Material  &  Class  \\  [0.5ex] 
\hline \hline
133 \ \   &   $H_2S$  (at $150$ $GPa$ pressure) &   Hydrogen-based superconductor \\ \hline
 110  \ \      &   $HgBa_2Ca_2Cu_3O_x(HBCCO)$   &      \\
93 \ \  &  $Bi_2Sr_2Ca_2Cu_3O_{10}(BSCCO)$ &  Copper-oxide superconductors   \\
55  \ \ &  $YBa_2Cu_3O_7 (YBCO)$ &  \\ \hline
41  \ \ &  $SmFeAs(O,F)$ &     \\
26  \ \ &  $CeFeAs(O,F)$ &   Iron-based superconductors  \\
18  \ \ &  $LaFeAs(O,F)$ &     \\ \hline
10  \ \ &  $Nb3Sn$ &     \\
9.2  \ \ &  $NbTi$ & Metallic low-temperature superconductors   \\
4.2  \ \ & $Nb$ &     \\
 [1ex] 
\hline 
\end{tabular}
\label{tb1} 
\end{table}

There have been many reports on justifying superconductivity phenomenon by condensed matter and many-body models, e.g. Fritz and Heinz two fluid QM theory model (1935), Ginzburg and Landau model (1950), and Cooper's BCS pair model (1975). Cooper's pair theory is not able to justify the high $T_c$-superconductivity \cite{1,2,3}. There have been several scientific reports in this field dealing with the basis of increasing the superconductivity transition temperature. In recent years, scientists have reached a transition temperature of about $140K$ in atmosphere pressure \cite{4}. However, reaching higher temperatures in normal pressure is a great challenge.

On the other hand, in magnetism theory, ordinarily, magnetic poles come in pairs: they have both a north pole and a south pole. However, a magnetic monopole is a magnetic particle having only a single, isolated pole or a north pole without a south pole, or vice versa. Although, at normal conditions of pressure and temperature, observation of a magnetic monopole has yet been confirmed, conversely with phenomenological model using the Maxwell's electromagnetic equations, existence of magnetic monopole with $Q_M =137e$ is presentable as following:
\begin{equation}
\nabla\cdot E=4\pi\rho_E, \ \  \ \nabla\cdot B=4\pi\rho_M,
\end{equation}
where Gaussian cgs units is used.  Here, $\rho_E$  and $\rho_M$are electrical and magnetic charge density. The nature of these monopoles with theory of quantum mechanics explored by Dirac in 1931 \cite{21}. Dirac introduced the magnetic monopole in order to explain the quantization of the
electric charge, which follows from the existence of at least one free magnetic charge.
He established the basic relationship between the elementary electric charge e and the basic
magnetic charge $g$:
\begin{equation}
eg=n\hbar c/2 \ \ \ \ \mbox{ Dirac quantization condition (DQC)   },
\end{equation}
where $n$ is an integer, $n = 1, 2, ...$.  The magnetic charge is $g = ng_D$; $g_D = \hbar c/2e = 68.5e$ is
called the unit Dirac charge.
The existence of magnetic charges and of magnetic currents
would symmetrize in form the Maxwell's equations, but the symmetry would not be perfect. After that, many activities were carried out to confirm it, so that, in 1959 by Y. Aharonov and D. Bohm experiment, the unit charge of magnetic monopole $Q_M= 137/2 e$, was obtained \cite{22}. In 1975, P. B. Price and E. K. Shirk investigated the detection of a moving magnetic monopole \cite{23}. They conducted an experiment and detected a very heavy particle. This heavy particle considered to be a magnetic monopole with electric charge of $Q_M=137e$ and mass of bigger than 200 proton mass ($m_p$). This experiment was in agreement with investigations of charged particles by V. R. Malkus in 1951 on arbitrary magnetic moment, moving simultaneously through the field of a magnetic monopole and an external electric field \cite{24}. It was concluded that the magnetic monopole can be coupled to matter with energies comparable to, but not significantly greater than, the material chemical bond. Investigations were done in the case of hydrogen where the lowest energy state depends upon the mass of the monopole.

In 1982, the first results from a superconductive detector for moving magnetic monopoles was reported by Blas Cabrera \cite{25}. Considering a magnetic charge moving at velocity $v$ along the axis of a superconducting wire ring with certain radius, the velocity - and mass-independent search for moving magnetic monopoles was performed by continuously monitoring the current in a $20-cm$ area superconducting loop. A single candidate event, consistent with one Dirac unit of magnetic charge, has been detected during five runs. However, with the development of researches in this field, heavy magnetic monopole initially suggested by 't Hooft and Polyakov in the framework of $SU (2)$ gauge theories \cite{3,4,5,25}.
In 2014, nearly 85 years after pioneering theoretical physicist P. Dirac predicted the possibility of their existence, an international collaboration was created, identified and photographed synthetic magnetic monopoles. Recently, prediction of magnetic monopoles in the spin-ice compounds such as $HO_2Ti_2O_7$ and $Dy_2Ti_2O_7$ has attracted much interest \cite{14}. Indeed, motion of magnetic monopole related to collective correlation of electrons with each other.
Our proposed models try to link the superconductivity to mobile magnetic monopoles which have major effect on the superconductive properties.

\section{Energy of magnetic monopole}\label{sec3}
\subsection{Electromagnetic semi-classical model and energy calculation of magnetic monopole}\label{a}
In this subsection, by utilizing electromagnetics and quantum physics approaches, electrostatic energy and thermal energy of magnetic monopoles are determined and possibility of existence of magnetic monopoles in temperature range of $T \leq 313 K$ is demonstrate.

\begin{figure}[ht]
 \includegraphics[height=4cm,width=8cm]{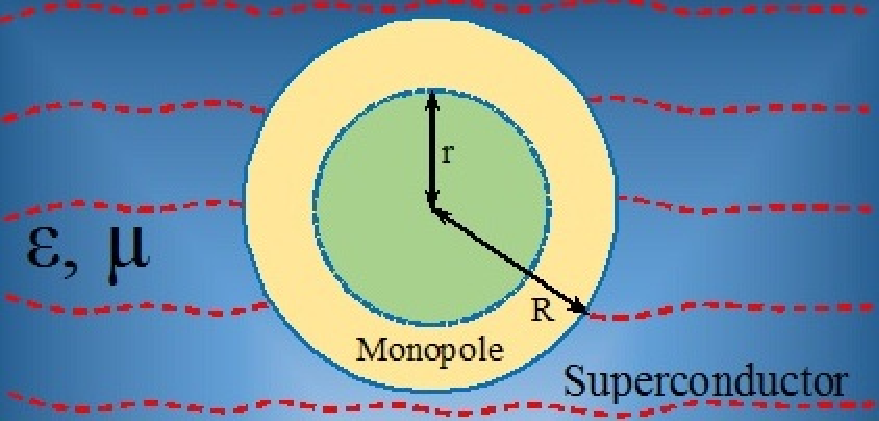}
  \caption{{A magnetic monopole as a charged sphere with uniform distribution of $Q$ and radius $R$
in superconductor medium ($\epsilon, \mu$).}}
\label{Fig1}
 \end{figure}

In this model, we assume that monopole exit in superconductor and consider a sphere with uniform charge distribution $Q$ and radius $R$ as an example of a magnetic monopole in superconductive medium
(see Fig. \ref{Fig1}).
 Then, we calculate the electrostatic energy of this sphere is as follows.
As the first step, we consider the electrical field ($E$) at distance $r$ from origin of charged sphere with radius $R$ and electrical permittivity ($\epsilon$) as following:
\begin{eqnarray}
\epsilon\int E\cdot dA =q_{net.enc} \Rightarrow E=\frac{Q}{4\pi\epsilon R^3} r.
\end{eqnarray}
Then, we compute the density of electrostatic energy in $r$ distance from origin of sphere as
\begin{eqnarray}
u=\frac{1}{2}\epsilon E^2 =\frac{Q^2}{32\pi^2\epsilon R^6}r^2. \label{4}
\end{eqnarray}
Therefore, the total electrostatic energy of sphere to radius $R$ and charge $Q$ is obtained as,
\begin{eqnarray}
U=\int udv =\int \frac{Q^2}{32\pi^2\epsilon R^6}r^2(4\pi r^2 dr)=\frac{Q^2}{40\pi\epsilon R}\label{5}
\end{eqnarray}

Now, we will continue by two different approaches to obtain "Total electrostatic energy of the  particle" with data for one typical matter.
\subsubsection{Semi-classical approach}
In this case, for one typical matter by considering eq. (\ref{4}) and using the magnetic
monopole with $Q=137 e$ (magnetic charge unit),   $R=(10)^{-7} m$  (size of cooper pair) and
$\epsilon \cong 2000 \epsilon_0$   \cite{1,2,3,4}, we can obtain the electrostatic energy
($U_E$) as:
\begin{eqnarray}
U_E \cong 2.16 \times 10^{-21} j \cong 1.35 \times 10^{-2} eV,\label{6}
\end{eqnarray}
where $e=1.6\times 10^{-19} C$, and $\epsilon_0=8.85\times 10^{-12} F/m$. Note the attainable values of $R$ and $\mu$ for this arbitrary matter (see TABLE \ref{tb2}).
\begin{table}[ht]
\caption{Relative permittivity of some materials at room temperature \cite{28}} \vspace{.3cm} 
\centering 
\begin{tabular}{c c  } 
 \hline\hline 
 Material    \ \  &  $\epsilon_r $   \\  [0.5ex] 
\hline \hline
Pyrex (Glass)\ \   &   $4.7 (3.7-10)$    \\ \hline
Diamond \ \      &   $5.5-10$         \\  \hline
Graphite \ \  &  $10-15$   \\   \hline
Silicon  \ \ &  $11.68$   \\ \hline
Glycerol  \ \ &  $42.5$ at $(25 {}^{\circ} C)$      \\ \hline
Titanium dioxide  \ \ &  $86-173$   \\   \hline
Strontium titanate \ \ &  $310$    \\ \hline
Barium strontium titanate \ \ &  $500$       \\  \hline
Barium titanate  \ \ &  $1200-10,000$ at $ (20-120  {}^{\circ}  C)$   \\  \hline
Lead zirconatetitanate \ \ & $500-6000$      \\  \hline
Conjugated polymers \ \ & $1.8-6$ up to $100,000$      \\  \hline
Calcium copper titanate \ \ & $>250,000$   \\
 [1ex] 
\hline 
\end{tabular}
\label{tb2} 
\end{table}

Now, we assume that the magnetic energy is equal to the electrostatic energy; and as a result the total energy ($U_{Total}$) can be obtained as:
\begin{eqnarray}
U_{Total} =U_E+U_B =2\times 2.16\times 10^{-21} j \cong 2\times 1.35 \times 10^{-2} eV.
\end{eqnarray}
On the other hand, thermal energy ($U_{Thermal}$) corresponding to this total energy ($U_{Total}$) is given by,
\begin{eqnarray}
U_{Thermal}= U_{Total}\cong 2\times 2.16 \times 10^{-21} j \cong 2\times 1.35 \times 10^{-2} eV \approx K_bT,
\end{eqnarray}
where $K_B =1.38 \times10^{-23} J/K$ is the Boltzmann constant. Indeed, $U_{Thermal}$   is the binding energy of magnetic monopole, i.e. if the thermal energy is greater than the above mentioned value, the magnetic monopole is broken. It means that the critical temperature ($T_c$) to form the monopole as a heavy particle in matter is
\begin{eqnarray}
T_c=\frac{U_{Thermal}}{K_b}
 =313K,
\end{eqnarray}
which is clearly about room temperature. We can find approximately similar result in quantum approach.
\subsubsection{Quantum approach }
In quantum mechanics with $Q=137/2  e$  and again $\epsilon \cong 2000 \epsilon_0$ one can obtain,
\begin{equation}
T_c=78 K. \label{10}
\end{equation}
 Therefore, by combining the quantum approach with classical approach for a magnetic monopole (having dielectric constant near $2000 \epsilon_0$), we conclude that the existence of magnetic monopole in temperature range of  $T \leq313K$ is possible.

Note that, these equations are valid for $N$ magnetic monopoles, because $N$-content is removed from both sides. Of course, this range depends on the dielectric constant, where increasing the dielectric constant ($\epsilon$) results in the elevation of the critical temperature ($T_c$).
\subsection{Energy calculation of magnetic monopole: Topological phase transition model}\label{b}
\begin{figure}[ht]
  \includegraphics[height=4cm,width=8cm]{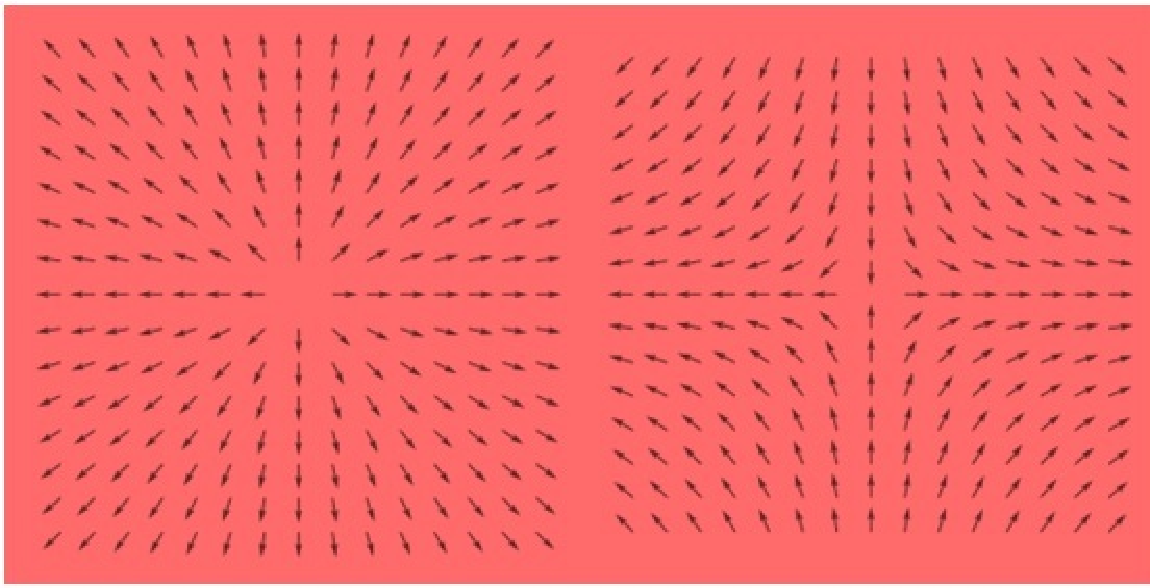}
  \caption{{Formation of vortex as a topologic defect; (left): vortex and (right): anti-vortex cite{18}.}}  \label{Fig2}
 \end{figure}

\begin{figure}[ht]
\begin{tabular}{c c c c}
\subfigure[]{\includegraphics[width=.5\linewidth]{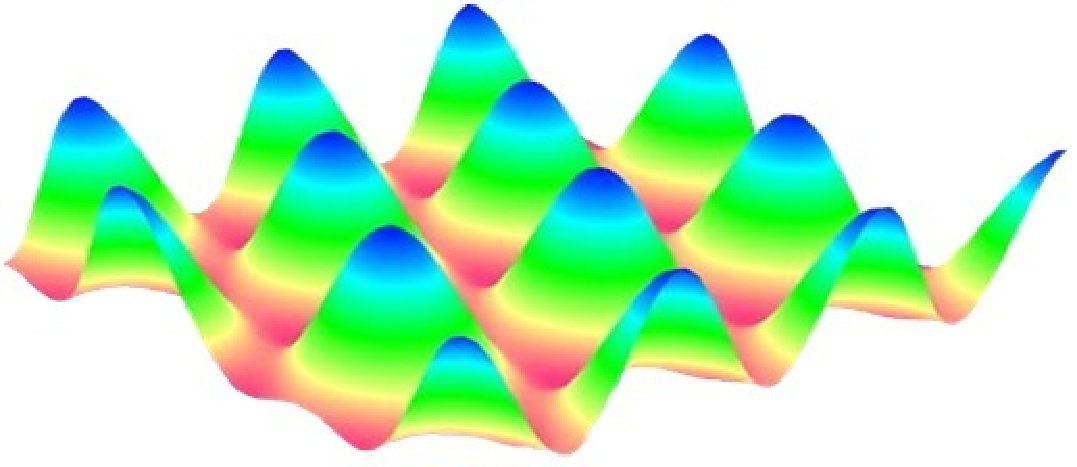} }\\
\subfigure[]{\includegraphics[width=.5\linewidth]{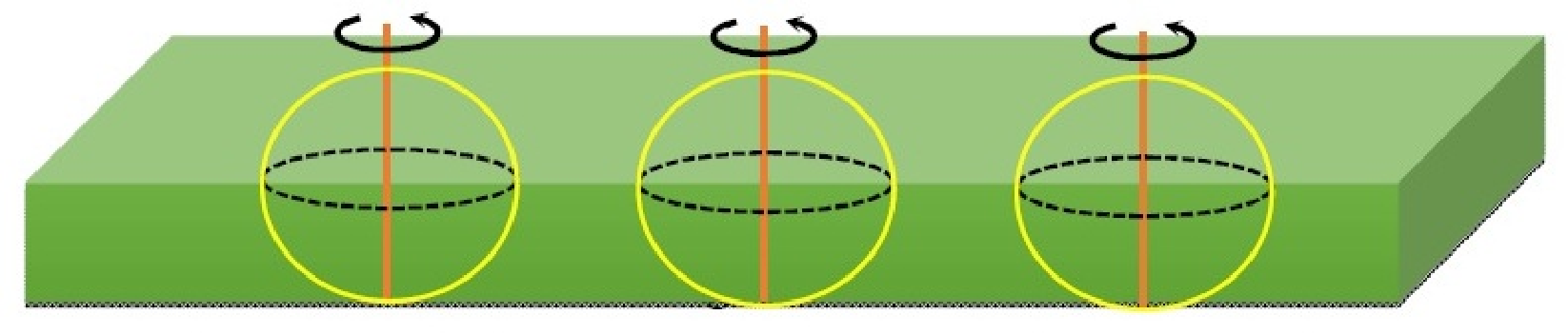}}\\
\subfigure[]{\includegraphics[width=.5\linewidth]{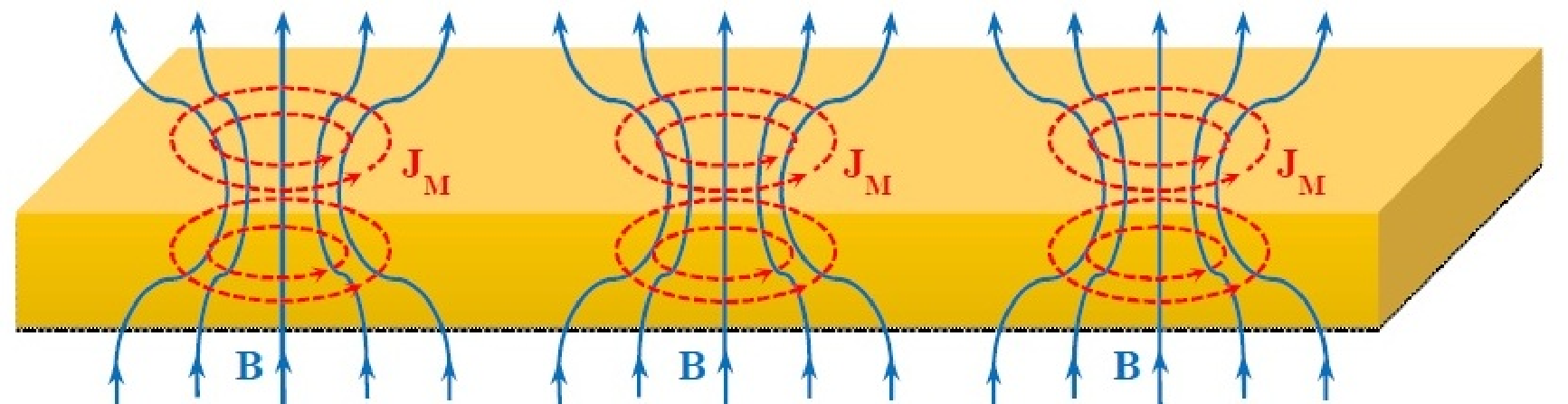}}
\end{tabular}
\caption{A schematic view of (a) the array of vortex waves and (b) magnetic monopole
systemsor "ball vortices" and (c) magnetic quantum flux related to magnetic monopolesin a
superconductor or "fluxon".}
\label{fig3}
\end{figure}

In this section, we are going to study the magnetic monopole model inspired by topological phase transition model \cite{19}. The TKH topological phase transition model introduces a vortex as a "hole" or "topologic defect" in matter. These vortices are responsible for electrical conductivity and critical phases in condensed matter (see figure \ref{Fig2}).
On the base of T-K topological phase transition, by quite a simple thermodynamic argument, the critical temperature for the phase of a single vortex is
       \begin{equation}
          T_C=\frac{\pi J}{2K_B},\label{11}
\end{equation}
where $K_B$ is Boltzmann constant, and constant $J$ determines the average strength of interaction between vortices \cite{19}.
Here, we postulate that the vortex is like a "magnetic monopole" in quasi-particle projection which simultaneously transmits and rotates around itself in matter and we can name it as "ball vortex" (figure \ref{fig3}).

The vortex has a size, a core and a "twisting mode" and naturally is a "Boson". That means ball vortex could be treated as a microscopic particle with anisotropy \cite{26}. The vortex in three dimensions shows a combination of spherical and chiral symmetry and very high anisotropy. On the other hand, in quantum approach, vortex is the quantization of flux which is called a "Fluxon".
Now, considering the breaking energy of a magnetic monopole or vortex as a topological defect at typical hot superconducting transition temperature, for example we can set $T_C=100 K $ (for a typical high temperature superconductor) in the equation (\ref{10}) to obtain constant $J$ as follow,
 \begin{equation}
          T \frac{\pi J}{2K_B} \Rightarrow  J =\frac{2K_B T }{\pi } \xrightarrow{\text{  T=100K}} J\cong\frac{276}{\pi}\times 10^{-23} j.
\end{equation}
   On the other hand, we can obtain energy of a single vortex as \cite{19}
   \begin{equation}
   U=\pi J\log\frac{r}{a},
   \end{equation}
where separation distance between vortices denoted by $r$, and $a$ is a short distance cutoff which can
be interpreted as the size of the vortex core. The energy of a single vortex is obtained as follows (J-
value form the equation (\ref{11}) and other experimental data),
\begin{equation}
   U= \pi J\log\frac{r}{a}\Rightarrow{\text{  r=1000nm; l=100nm }} U\cong 640\times10^{-23} j \cong
   4\times10^{-2} eV,
   \end{equation}
   where we set $r=10^3$  $nm$ and $a=10^2$ $ nm$.
This value of energy is in the range of the electromagnetic energy of a magnetic monopole with
  $Q=137 e$, which we obtained before in the electromagnetic model in section \ref{sec3} (relations
  (\ref{5}) and (\ref{6})).

Therefore, we conclude that the magnetic monopole model is in agreement with results of topological TKH
phase transition model.
\subsection{Eigenvalue calculation of grand state of magnetic monopole: Solid state model}\label{c}
In this model, we assume that monopole moves in periodic lattice of a solid with effective mass of $m^*$. We also calculate the expectation value of Hamiltonian <H>with wave function of magnetic monopole similar to particle in box with quantum dimensionsfrom the following relation:
\begin{eqnarray}
E_k = \langle H\rangle =\frac{\int_{Box} d^3x \psi^*H\psi}{\int_{Box} d^3x \psi^*\psi},
\end{eqnarray}
$H=\frac{p^2}{2m^*}$ and $\psi(x)= e^{i\lambda+ikx}$ (wave function of free magnetic monopole in
 $x$-dimension) \cite{21}. Here, we ignore the monopole-monopole interaction and potential energy in approximation of in-depended monopole. This leads
\begin{eqnarray}
E_k   =\frac{\int_{Box} d^3x e^{-i\lambda-ikx}\frac{1}{2m^*}\left(\frac{\hbar}{i}\right)^2\frac{d^2}{dx^2} (e^{i\lambda+ikx})}{\int_{Box} d^3x e^{-i\lambda-ikx}e^{i\lambda+ikx}} =\frac{\hbar^2k^2}{2m^*}.
\label{16}
\end{eqnarray}
According to $k=2\pi/a$ (i.e. dimension of box or unit cell so that   $- \pi/a+ \pi/a$  in Brillouin Zone), Relation (\ref{16}) leads to
\begin{eqnarray}
E_k     =\frac{h^2}{2m^*a^2}.
\end{eqnarray}
Here, $m^*$ is effective mass of magnetic monopole which is $4700\ m_e$for "light monopole", based on the Dirac's formulation, monopoles are assumed to exist as point-like particles \cite{27} and $ m_e = 9.11\times 10^{-31}\ Kg$.
For example, in a typical lattice with dimension ($a$) of box or unit cell ($a=1\times 10^{-9}\ m$),
$h = 6.63\times 10^{-34}$ $js$, therefore for $3$-dimension, eigenvalue:
\begin{equation}
E_k=1.54\times10^{-22}\ J \simeq 10^{-3}\ eV.
\end{equation}
According to the equipartition theory of energy, $E_{Total}= \langle H_k\rangle + \langle U\rangle$, so if expectation value of potential energy added to kinetic energy,total energy is in the range of last models.
\subsection{Dynamic model of magnetic monopole motion monopole}\label{d}
Understanding the motion dynamics of magnetic monopole helps study of the role of the relaxation time and resistivity in superconductors.
It is known that that dc electrical conductivity in low-dimension depends on charge density ($n$), relaxation time ($\tau$ ) and mass ($m$) according to the following equation:
\begin{eqnarray}
\sigma=\frac{nq^2\tau}{m},
\end{eqnarray}
which is obtained from the "Drude" classical model in solid state physics \cite{28}.

Accordingly, in the presented model because of the low number of magnetic monopoles ($n\sim 10^4$  $cm^{-3}$), conductivity highly depends on relaxation time which is defined as duration of collision between two particles. Indeed, the low number of magnetic monopole carriers, lead to relaxation time increases extremely. For example, in comparison with silver, with electron density ($n = 5.86\times 10^{22}$) in $300K$ and relaxation time ($\tau  = 4\times 10^{-14}$ $s$) \cite{28}; if $\tau$  increases to $\sim 10^{-4}$ $s$ (a reasonable level), the conductivity increases about $10^{10}$ times which can be considered close to superconductivity conditions.

Moreover, it should be noted that the origin of the electrical resistivity in normal state of solids is the high concentration of charge carriers (electron) with very low mass which makes the rate of random collisions very high, similar to a classical gas. There are also other factors including structural properties such as lattice defects, grain boundaries and arrangement of atoms in building of matter. All these are effective when meeting very small particles (such as electrons). But if a massive and large particle similar to magnetic monopole is the source of conductivity, not only these factors are unimportant but also can cause the heavy particle transmit and the solid lattice acts as a background for speeding up the movement in the lattice.Indeed, periodic lattice of solid acts as a mesh with very small apertures, so that a heavy particle can move forward on it with getting help from this background without preventive.

Therefore, we can consider this phenomenon like movement of a heavy ball on a surface that the very fine cracks and friction help its rotation and movement, as if this friction disappears, movement is disturbed.

These conditions can expand to superconductor medium and motion of magnetic monopole as heavy particle on periodic lattice among plates under an external field. Therefore, here, we can conclude that electrical resistivity which is caused by structural effects becomes insignificant (i.e  $ \rho \rightarrow 0$).

Considering the above argument, the total dynamic energy of the magnetic monopole as following:
\begin{equation}
E_{Total}=E_t+E_r+E_\nu=\frac{1}{2}mv^2+\frac{1}{2}I\omega_r^2+\left(n+\frac{1}{2}\right)h\omega_\nu,
\end{equation}
where $E_t$, $E_r$  and $E_v$   are kinetic energy, rotation energy and Debye quantum vibration energy terms, respectively.

For example, in a matterby substituting $M = 200\ m_p$, $v = 100\ m/s$ \cite{28}, $I = 2/5 MR^2$, $R = 100\ nm$, $n= 0$ (ground state) and $\omega_r$ and $\omega_\nu$ are ordinary, then we can obtain:
\begin{eqnarray}
E_t&=&1.67\times 10^{-21}\ J =1.04 \times 10^{-2}\ eV, \\
E_r&=&6.68\times 10^{-40}\omega_r^2\ J =4.18 \times 10^{-21}\omega_r^2\ eV,\\
E_\nu &=&3.31\times 10^{-34}\omega_\nu^2\ J =2.07 \times 10^{-15}\omega_\nu^2\ eV.
\end{eqnarray}
As seen, terms $E_r$ and $E_\nu$ are very small (with any $\omega_r$ and $\omega_\nu$) in comparison with kinetic energy $E_k$. Hence, we can write total energy just according to kinetic energy:
\begin{equation}
E_{Total}=1.04 \times 10^{-2}\ eV.
\end{equation}
This energy is equal to electromagnetic energy and vortex energy as obtained previously.
Of course, in these  models, it is important to note that the optimum amount of magnetic charge i.e. $Q=137e$ is necessary, since as the charge increases, the amount of monopole mass increases and mobility decreases.

\section{Discussions}\label{sec4}
\begin{figure}[h!] \label{Fig4}
 \begin{center}
 {
 \includegraphics[height=8cm,width=5cm]{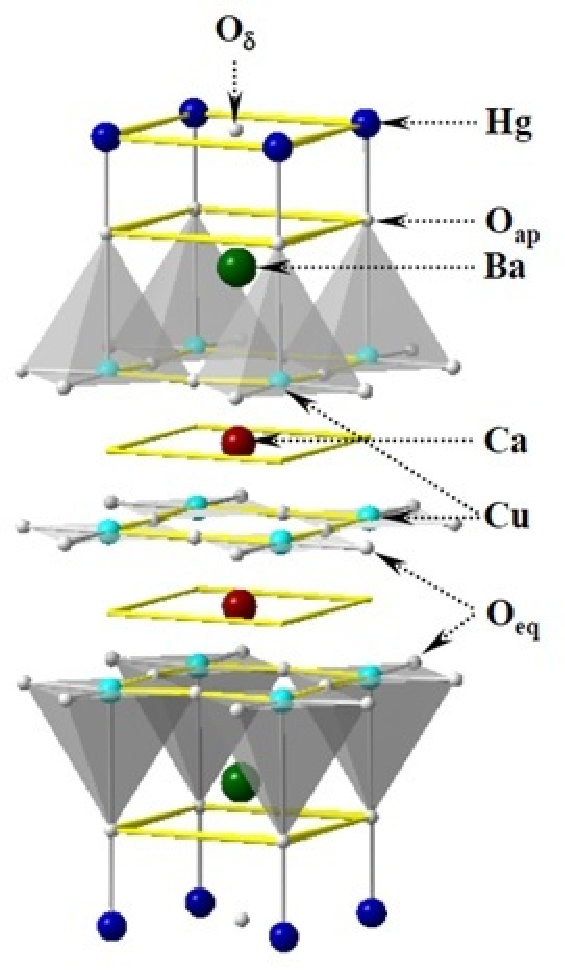}
  \caption{\small{Crystal structure of mercury compound $Hg-1223$ as a high $T_c$ superconductor with
layered structure \cite{29}.}}}
 \end{center}
 \end{figure}

In this section, with reference to physical arguments and previous scientific reports, we are going to correlate superconductivity and existence of magnetic monopoles by investigating the structural parameters of the material.
So far, it has been reported that the superconductivity properties of high-$T_c$ superconductors are determined by \cite{29, 30}
\begin{itemize}
\item type of layered crystal structure and topological defects (or vortices),
\item bond energy and lengths and space between separated layers,
\item valence and mass of the ions,
\item Coulomb coupling between neighboring electronic bands,
\item volume of primitive cell and internal pressure (or strong strain) in lattice structure.
\end{itemize}
Therefore, the mentioned parameters play a key role in superconductivity behavior of the material.
On the other hand, high permittivity is related to strong strain in the lattice and configuration of compound structures. Therefore, above mentioned parameters can also play a key role in dielectric behavior of matter. Here, we are going to propose that the magnetic monopole under very high interlayer pressure and high dielectric constant can be the responsible potential for High-$T_c$ superconductivity. The following justifications can help understanding this connection.
Attention of a number of researchers has been drawn to the fact that almost all high-$T_c$ superconductors are layered compounds. The layered structure introduces anisotropy to the material behavior both in the normal and in the superconducting states. This is found by perovskites group in high-$T_c$ superconductors, that the superconducting state is created by substituting ions of different valences in layers \cite{30}. Indeed, recent developments in the field of material science have opened the way to engineer the dielectric response function of superconducting materials on the nanometer scale in order to increase the superconducting critical temperature. For example, $HgBa_2Ca_2Cu_3O_{8+\delta}$ and $\kappa-(ET)_2Cu[N(CN)_2] Cl$ become superconductors at normal pressure and $25 GPa$ and $30 MPa$. Here, special effects of the lattice strain and dielectric constant lead to ultra-high pressure in sub-lattice \cite{31, 32}. The polymer compound of $\kappa-(ET)_2Cu[N(CN)_2] Cl$, where their crystal structure consists of alternating layers of ET molecules ($C_10S_8H_8$) and insulating anion sheets, is an important organic superconductor. The crystal structure of mercury compound $HgBa_2Ca_2Cu_3O_{8+\delta}$ ($Hg-1223$) with three copper layers is shown in Fig. \ref{Fig4}.

Under chemical pressure in $Hg-1223$, the $Cu-O-Cu$ bond in the plane does not change, while application of an external pressure causes the buckling of the $Cu_2O$ plane. The structural instability and the largest copper-apex oxygen distance have significant influence on the highest $T_c$ for the mercury compounds.
On the other hand, ultra-high internal pressures in order to $10^4$ bar in sub-lattice of layered structures such as "perovskites" with complex structure lead to formation of structural "holes" or "vortices" as topological defects. Therefore, mass of ions, room pressure, vacancy and composition of layers, and low dimension, play an important role in studying high-temperature superconductors.
 Finally, according to the reliable evidences, existence of magnetic monopole at very high pressure in order of G-Pascal at free space is confirmed, and high temperature superconductivity at very high pressure G-Pascal up to $Tc\sim 200K$ has been obtained \cite{31}.  Therefore, we predict that in ultra-high internal pressure and high condense state of matter in normal state, magnetic dipole split from each other and magnetic monopole is formed inside layered solids and under this condition hot superconductivity is available. These two phenomena: layered perovskite structure and temperature transition is related with each other.

\section{Conclusion}\label{sec5}
\begin{figure}[ht]
 {
 \includegraphics[height=4cm,width=8cm]{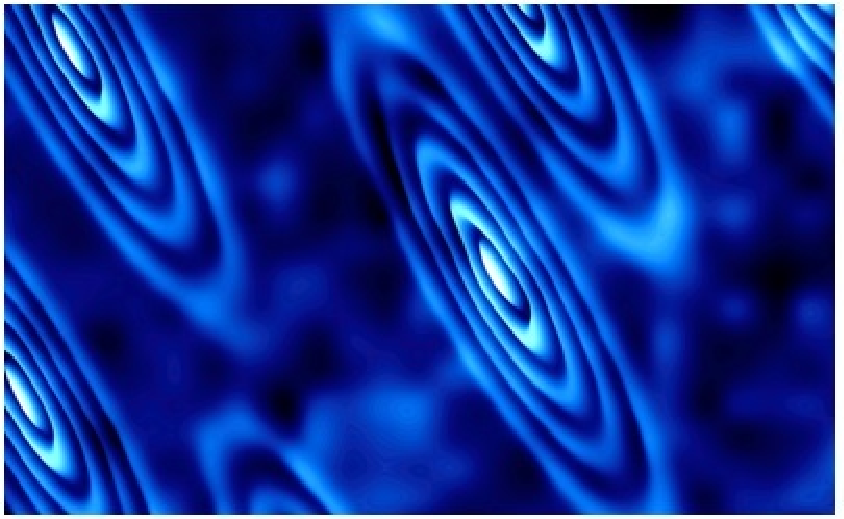}

  \caption{Observation of a "Nematic Hall Quantum liquid" on the Surface of Bismuth thin film
by STM in October 21, 2016 \cite{32}.}}
\label{Fig5}
 \end{figure}
In this paper, we have presented the simple models, with three different physical approaches, for establishing a relation between the superconductivity and the mobile magnetic monopoles in the superconducting medium.
Our results from linking the magnetic monopoles and superconductive properties of material is summarized as follows.
\begin{itemize}
\item Superconductivity at high transition temperatures is highly depend on high dielectric constant ($\epsilon >1000$). By increasing the dielectric constant, the transition temperature $T_c$ is elevated (see section \ref{a}).
\item Superconductivity at high temperatures even close to room temperature is possible (section \ref{a}).
\item Existence of magnetic monopole is dependent on ultra- high internal pressure, which takes place inside   a super lattice solid with high dielectric constant and complex structure (section\ref{sec3} and \ref{sec4}).
\item Origin of the superconductivity may be magnetic monopoles (section \ref{sec3} and \ref{sec4}).
\item The electromagnetic model of monopole (section 3(a)) matches with TKH topological model (section \ref{b}).
\item The motion of magnetic monopole with $Q=137 e$ and mass $> m_p$ matches vortex as a topological defect (section \ref{b} and \ref{d}).
\item  The dynamic total energy of magnetic monopole is agreement with vortex energy (topological model) and electromagnetic energy of charged particle with $Q=137 e$ or $Q=137/2 e$ in classic and quantum approaches, respectively (section \ref{sec3}).
\item The ultra-high internal pressures in order to $10^4$ bar in sub- lattice of layered structures such as "perovskites" with complex structure is very effective for formation of structural "holes" or "vortices" as topological defects which are single magnetic monopoles (section \ref{sec4}).
\end{itemize}
However, development of these models are necessary as advanced theoretical methods by considering other topics such as; nature of the coupling mediation of monopoles, dependence on monopole effective mass, scattering and structural topological effects.
For detection of magnetic monopole, we propose experimental techniques under gradual conditions from high temperature to low few temperatures, especially for layered structures like to High-$T_c$ perovskite superconductors by advanced scanning tunneling microscope (STM) similar to Fig. \ref{Fig5} for observation of a "Nematic Hall Quantum liquid" on the surface of bismuth thin film by E. Benjamin and et. al \cite{33}.
 In other hypothetical experiments, with the temperature dropping, we can observe the accumulation of magnetic lines around a magnet on the one pole.

\end{document}